\documentclass[12pt]{iopart}
\usepackage{epsfig}
\usepackage{graphics}
\usepackage{amstext}

\begin{document}
\title[]{GravEn: Software for the simulation of gravitational wave detector network response}
%\title[]{Overview of a Robust Software Gravitational Wave Simulation Generator}

\author{Amber L. Stuver, Lee Samuel Finn}
\address{Center for Gravitational Wave Physics, The Pennsylvania State University, University Park, PA 16802, USA}

\ead{\mailto{stuver@gravity.psu.edu}}

\begin{abstract}

Physically motivated gravitational wave signals are needed in order to study the behavior and efficacy of different data analysis methods seeking their detection.  GravEn, short for Gravitational-wave Engine, is a MATLAB$^{\textregistered}$\ software package that simulates the sampled response of a gravitational wave detector to incident gravitational waves. Incident waves can be specified in a data file or chosen from among a group of pre-programmed types commonly used for establishing the detection efficiency of analysis methods used for LIGO data analysis.  Every aspect of a desired signal can be specified, such as start time of the simulation (including inter-sample start times), wave amplitude, source orientation to line-of-sight, location of the source on the sky, etc.  Supported interferometric detectors include LIGO, GEO, VIRGO and TAMA.

\end{abstract}


\section{Introduction}

Gravitational-wave Engine, or GravEn, is a compilable MATLAB$^{\textregistered}$~\cite{MATLAB} software package developed to simulate the output of a network of gravitational wave detectors to incident gravitational waves. It is used within the LIGO Scientific Collaboration to create the simulations used in the so-called ``Mock-Data Challenges,'' or MDCs, to determine the efficiency of the detection pipeline for gravitational wave bursts. It can be used for Monte Carlo studies of sensitivity to source distributions that vary in spatial density, intrinsic energy, or other parameters. The source, together with full operational documentation~\cite{GravEnMAN}, is available for download from~\cite{GravEnCVS}.

Every signal generated by GravEn is associated with a ``source'' that is located at a particular point on the celestial sphere, has a particular orientation with respect to the line-of-sight to the detector, and whose signal arrives at Earth's center at a particular moment of time. The source may be astrophysical in origin, in which case the waveform and orientation have a physical motivation, or it may be ad hoc, in which case the ``line-of-sight'' is just the propagation direction of a plane wave and the ``orientation'' is related to the basis on which the waveform in the two polarizations is specified. From the waveforms in the two polarizations, the line-of-sight, orientation, time-of-arrival information, and each detector's location, orientation, and response function, GravEn determines when waves are incident on each detector on Earth's surface and the sampled response of each detector to the incident waves.

Simulated waveforms can be generated from within GravEn or waveforms can be provided as external input from a file.  The use of pre-generated waveforms allows for the inclusion of astrophysically motivated waveforms from external numerical projects and the rapid evaluation of data analysis methods on these state-of-the-art numerical simulations.

All calculations are performed in a coordinate system with the origin located at the center of the Earth.  The time of flight between the center of the Earth and the detector is computed knowing the source's location on the sky and the start time of each simulation is then adjusted accordingly.  This allows for the coincident (within time-of-flight) generation of simulations between any set of detectors.  GravEn currently supports the LIGO, GEO, VIRGO and TAMA interferometric gravitational wave detectors and the addition of other detectors at other locations is straightforward.

The general operation of GravEn is summarized in Section \ref{summary} followed by a discussion of the major physical components of GravEn and their infrastructure.  These include the time projection between the center of the Earth and detector (Section \ref{timedelay}), metric perturbation generation (Section \ref{makeh}), TT gauge projection (Section \ref{makett}), and detector projection (Section \ref{detproj}).  Finally, this overview concludes with a commentary on injecting software simulation signals into actual detector data in Section \ref{calib}.

\section{Driver Function - \texttt{graven}}

\subsection{Structure of the Driver} \label{summary}

There are four modes of operation for GravEn: 1) information describing the simulation is read from a file, 2) information describing a \emph{distribution} of sources is provided in a file, and simulations are drawn randomly from the distribution, 3) the driver takes the simulation information directly from the command line and does not need to read a file and 4) a pre-generated waveform is pushed through the driver.  In the first three cases waveforms are selected from a list provided by GravEn while the last takes its waveform data from an external source.

\subsubsection{Generate Original Waveforms}

GravEn determines its mode of operation and, if necessary, the names of any input files, from its command line input. From these it generates a master list of simulations that it will perform. It then loops over this list, performing each simulation on the list in turn.

A simulation begins by determining, from the source location on the sky and the detector location on Earth, when the initial wavefront is incident on the detector. This is provided in in units of `samples' relative to when the initial wavefront is incident on Earth's center. These calculations are carried-out by the GravEn function \texttt{ifodelay}\footnotemark. One sample is the interval between samples at the output of the specified detector. Fractional samples are permitted. 

\footnotetext{Function names are specified in \texttt{fixed width font} throughout this work.}

The second step in the simulation process is the determination of the TT gauge metric perturbation incident at Earth's center and, from it and the time of arrival of the wave at the detector, the projection of the wave on the geometric part of the detector's response function\footnotemark. \footnotetext{Since GravEn simulates detectors in the short-antenna approximation, the detectors response can be decomposed into a frequency-independent piece, which is determined by the detectors geometry, and a frequency dependent piece, which is determined by the detector's principle of operation and its transducer electronics.}These calculations are carried out by the GravEn functions \texttt{makeh} and \texttt{makeTT}. 

The final step of the simulation process is the conversion of the geometrically projected strain into the sampled output of the detector transducer (e.g., volts), which may be added to or substituted for detector output as it is provided to a data analysis pipeline. The result is saved to an output file and all the information required to reproduce the simulation is saved to a log file, which GravEn can read directly to reproduce the output.

\subsubsection{Use Pre-generated Waveforms} \label{preGen}

The structure of the driver is exactly the same for pre-generated waveforms, except that once the pre-generated waveform is read from a file, the metric perturbation creation function, \texttt{makeh}, and the TT gauge projector, \texttt{makett} are bypassed.  Inter-sample start times are not yet supported for this mode of operation; so, the moment of incidence of the initial wavefront on the target detector is also rounded to the nearest sample.

Several assumptions are made about the pre-generated waveform:
\begin{itemize}
        \item The waveform is in the TT gauge
	\item The waveform is evenly sampled at the same rate as the detector\footnotemark
\end{itemize}

\footnotetext{Waveforms arising from large scale numerical simulations are generally not evenly sampled, e.g., \cite{Ottetal}; so, resampling is necessary to use these waveforms directly as input to GravEn.}

It is important to make sure that the sampling frequency input into GravEn matches that of the pre-generated waveform and the detector for accurate results, however GravEn does take into account the different sampling rate of VIRGO (20 kHz) versus the other interferometric detectors (16,384 Hz).

\subsection{Outputs}

The output of \texttt{graven} consists of the name of the simulation, the start time of the simulation in whole samples at the detector, the simulation time series, the start time of the simulation in seconds and nanoseconds at the detector, and the time of maximum strain in seconds and nanoseconds at the detector.  This information, along with the start time of the simulation at the center of the Earth, the maximum waveform amplitude and other source orientation input variables are recorded in a log file that can be used as an input file to generate coincident simulations between detectors or recreate simulations for a single detector.

\section{Time Projector Between the Center of the Earth and the Detector - \texttt{ifodelay}} \label{timedelay}

GravEn supports the generation of coincident simulations between multiple detectors by defining the time of arrival of the gravitational wave at the center of the Earth.  Then, for each detector, the time lapse between the arrival at the center of the Earth and the arrival at the location of the detector is calculated.  This lapse is added to the arrival at the center of the Earth and recorded as the appropriate signal arrival time for the specified detector. 

The time lapse is computed from the location of the detector on the surface of the Earth ($\vec{\mathbf{R}}_{\mathrm{det}}$), the unit vector of the sky location of the source ($\vec{\mathbf{n}}_{\mathrm{GW}}$) which is opposite to the direction of propagation, and the speed at which gravity propagates (i.e. the speed of light, $c$):
\begin{equation}
t=\frac{\vec{\mathbf{R}}_{\mathrm{det}}\cdot \vec{\mathbf{n}}_{\mathrm{GW}}}{c}
\end{equation}

\begin{figure}
\centering
\includegraphics[width=1.0\textwidth]{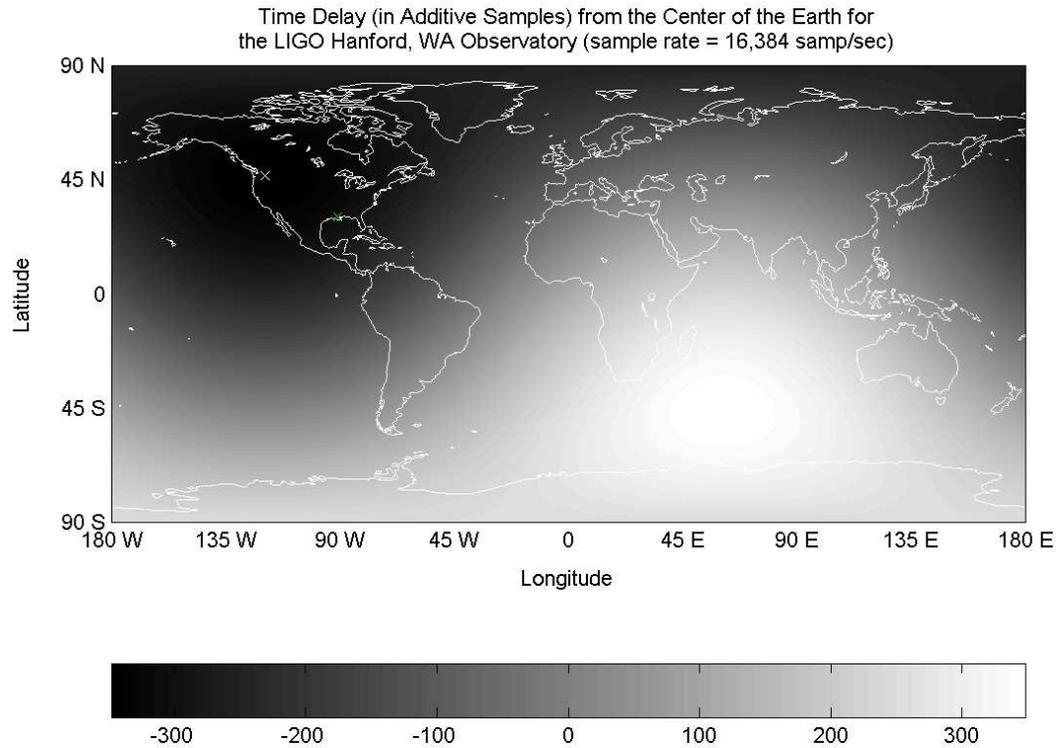}
\caption{The number of samples to add to signal start time with respect to the source sky location for the LIGO Hanford Observatory (sampling frequency=16384 samples/sec)}
\label{delayLHO}
\end{figure}

An example time projection for the LIGO Hanford Observatory, WA is shown in Figure \ref{delayLHO}.

\section{Metric Perturbation Function - \texttt{makeh}} \label{makeh}

The LIGO Scientific Collaboration uses several different styles of \emph{linearly-polarized} waveforms for the purpose of evaluating the efficiency of its gravitational wave analysis pipelines. GravEn supports the automatic generation of all these. The time-dependence of the waveforms that GravEn can generate internally are listed below:
\begin{eqnarray*}
A_{\text{G}}(t|f,\tau)&=&\exp\left(-\frac{t}{\tau}\right)^{2}\\
A_{\text{SG}}(t|f,\tau)&=&\sin(2\pi f t)\exp\left(-(t/\tau)^2\right)\\
A_{\text{CG}}(t|f,\tau)&=&\cos(2\pi f t)\exp\left(-(t/\tau)^2\right) \\
A_{\text{SINC}}(t|f,\tau)&=&\frac{\sin(\pi ft)}{\pi ft}
\end{eqnarray*}
as well as sine waves, cosine waves, linear chirps\footnotemark, Gaussian modulated linear chirps, square waves, Gaussian modulated square waves, sawtooth waves, Gaussian modulated sawtooth waves and Gaussian modulated white noise can be generated similarly to the ones described above. \footnotetext{The linear chirp begins at 0 Hz and sweeps up to the frequency specified.}Corresponding to linearly polarized character, GravEn postulates for each an axisymmetric source whose axis is taken to define the $Z$-axis of the \emph{source frame}. The metric perturbation in the source frame is taken to be of the form
\begin{equation}
h{\ldots}(t) = \frac{A_0}{r} A_{\ldots}(t)e_{\text{src}}
\end{equation}
where
\begin{equation}
e_{\text{src}} = \left(\begin{array}{rrr} 1&0&0\\ 
0&1&0\\ 
0&0&-2 
\end{array}\right)
\end{equation}

\section{TT Gauge Projector - \texttt{makett}} \label{makett}

The routine \texttt{makeh} returns the metric perturbation as a matrix-valued time series. The metric perturbation is expressed in the Cartesian coordinates of the source frame. The routine \texttt{makett} determines the TT gauge radiation incident at Earth's center in a frame related to Earth's orientation in space and the propagation direction of the radiation. 

Let $XYZ$ denote the Cartesian coordinate directions of the source frame, $xyz$ the Cartesian coordinate directions of an Earth-centered coordinate frame, and $\alpha\beta\gamma$ the coordinate directions of the radiation propagation frame:
\begin{itemize} 
\item Let the Earth frame have its origin at Earth's center with the $z$-axis aligned with Earth's angular momentum; 
\item Choose the $x$ axis to lie in Earth's equatorial plane, with the source in the positive $x$ half of the $xz$ plane; 
\item Choose the $y$ axis so that $\hat{x}\times \hat{y} = \hat{z}$; 
\item Let the $\gamma$ direction be in the direction of propagation of the radiation; 
\item Choose the $\alpha$ direction to be parallel to Earth's equatorial plane; 
\item Choose the $\beta$ direction so that $\hat{\alpha}\times\hat{\beta}=\hat{\gamma}$; 
\item Let the unit vector along the line-of-sight from the source to the detector be $\vec{n}$ in the Source frame; 
\item Let $\pi-\theta_{int}$ be the altitude angle of $\vec{n}$ relative to the source $XY$ plane; 
\item Let $\varphi_{int}$ be the azimuth angle of $\vec{n}$ relative to $x$ in the source frame; 
\item Let $\psi$ be the angle between the projection of $Z$ onto the $\alpha\beta$ plane and $z$ onto the $\alpha\beta$ plane, q.v. Section \ref{psi} and Figure \ref{psi3}. 
\end{itemize} 

With these definitions we can define the rotation that takes $XYZ$ into $\alpha\beta\gamma$: 

\begin{equation}
\left( \begin{array}{c}
\ \hat{\alpha} \\
\ \hat{\beta} \\
\ \hat{\gamma} \\
\end{array} \right) =
\left( \begin{array} {ccc}
\ \cos(\varphi_{int})\cos(\theta_{int}) & \sin(\varphi_{int})\cos(\theta_{int}) & -\sin(\theta_{int}) \\
\ -\sin(\varphi_{int}) & \cos(\varphi_{int}) & 0 \\
\ \cos(\varphi_{int})\sin(\theta_{int}) & \sin(\varphi_{int})\sin(\theta_{int}) & \cos(\theta_{int}) \\
\end{array} \right)
\left( \begin{array} {c}
\ \hat{X} \\
\ \hat{Y} \\
\ \hat{Z} \\
\end{array} \right) \\
\end{equation}

Using this rotation matrix we can re-express the metric perturbation $h$ in the radiation propagation frame. In the radiation propagation frame natural polarization projection tensors $\hat{e}_{+}$ and $\hat{e}_{\times}$ take on a particularly simple form
\begin{eqnarray} 
\hat{e}_{+} &=& \hat{\alpha}\otimes\hat{\alpha} - \hat{\beta}\otimes\hat{\beta}\\ 
\hat{e}_{\times} &=& \hat{\alpha}\otimes\hat{\beta} + \hat{\beta}\otimes\hat{\alpha} 
\end{eqnarray} 
Finally, we identify the radiation in the $+$ and $\times$ polarization states to be:
\begin{eqnarray}
h_{+}=\frac{1}{2}h^{ij}\hat{e}_{+ \ ij} \\
h_{\times} = \frac{1}{2}h^{ij}\hat{e}_{\times \ ij}
\end{eqnarray}

The function \texttt{makett} accepts as input $h$ as returned by \texttt{makeh}, the propagation direction of the radiation expressed in the Earth frame, and the angles $\theta$ and $\phi$, and returns $h_+$ and $h_{\times}$.

\section{Detector Projector - \texttt{detproj}} \label{detproj}

Given the output of \texttt{makett} (or a pre-generated waveform), a detector and the orientation of the source on the sky ($\theta_{ext}$, $\varphi_{ext}$, $\psi$), \texttt{detproj} projects the TT gauge waveform onto the antenna pattern of the detector.  $\theta_{ext}$ is the source's co-declination sky location as projected onto Earth's fixed coordinates, $\varphi_{ext}$ is the source's sky location longitude as projected onto Earth's fixed coordinates (in radians) and $\psi$ is the source's polarization angle in radians, defined in section \ref{psi}.  The detector is assumed to be the origin of a coordinate system with $\hat{Z}$ pointing toward the North Pole, $\hat{X}$ pointing toward the intersection of the Prime Meridian and the Equator, and $\hat{Y}=\hat{Z}\times\hat{X}$ and the gravitational wave is assumed to be a plane wave. 

A rotation of Earth's axes is needed so that $\hat{Z}$ points towards the source along the line-of-sight:
\begin{equation}
\left( \begin{array}{c}
\ \hat{e}_{x} \\
\ \hat{e}_{y} \\
\ \hat{n} \\
\end{array} \right) =
\left( \begin{array} {ccc}
\ \cos(\varphi_{ext})\cos(\theta_{ext}) & \sin(\varphi_{ext})\cos(\theta_{ext}) & -\sin(\theta_{ext}) \\
\ -\sin(\varphi_{ext}) & \cos(\varphi_{ext}) & 0 \\
\ \cos(\varphi_{ext})\sin(\theta_{ext}) & \sin(\varphi_{ext})\sin(\theta_{ext}) & \cos(\theta_{ext}) \\
\end{array} \right)
\left( \begin{array} {c}
\ \hat{X} \\
\ \hat{Y} \\
\ \hat{Z} \\
\end{array} \right) \\
\end{equation}

Since both the source's and Earth's Z axes now lie on the line-of-sight, we can define the + and $\times$ polarization basis in the source frame:
\begin{eqnarray}
\hat{e}_{+}=\hat{e}_{x}\otimes \hat{e}_{x}-\hat{e}_{y}\otimes \hat{e}_{y} \\
\hat{e}_{\times}=\hat{e}_{x}\otimes \hat{e}_{y}+\hat{e}_{y}\otimes \hat{e}_{x}
\end{eqnarray}
%\end{subequations}

Now define the + and $\times$ polarization basis in the detector frame:
%\begin{subequations}
\begin{eqnarray}
\hat{e}^{\prime}_{+}=\cos(2\psi)\hat{e}_{+}+\sin(2\psi)\hat{e}_{\times} \\
\hat{e}^{\prime}_{\times}=\cos(2\psi)\hat{e}_{\times}-\sin(2\psi)\hat{e}_{+}
\end{eqnarray}
%\end{subequations}
%
where $\psi$ is the polarization angle as described in Section \ref{psi}.

The antenna projection matrix is formed by:
\begin{equation}
	D=\hat{V}\otimes \hat{V}-\hat{W}\otimes \hat{W}
\end{equation}
where $\hat{V}$ and $\hat{W}$ are the unit vectors for the X and Y arms of the detector in the WGS-84 coordinate system.

The beam pattern functions are defined to be:
\begin{eqnarray}
        F_{+}=\frac{1}{2}D^{ij}\hat{e}^{\prime}_{+ \ ij} \\
	F_{\times}=\frac{1}{2}D^{ij}\hat{e}^{\prime}_{\times \ ij}
\end{eqnarray}

The polarization averaged antenna pattern is defined to be \cite{Tinto&Schutz, 300yrs}:
\begin{equation}
\rho^{2}=F_{+}^{2}+F_{\times}^{2}
\end{equation} 

\begin{figure}
\centering
\includegraphics[width=1.0\textwidth]{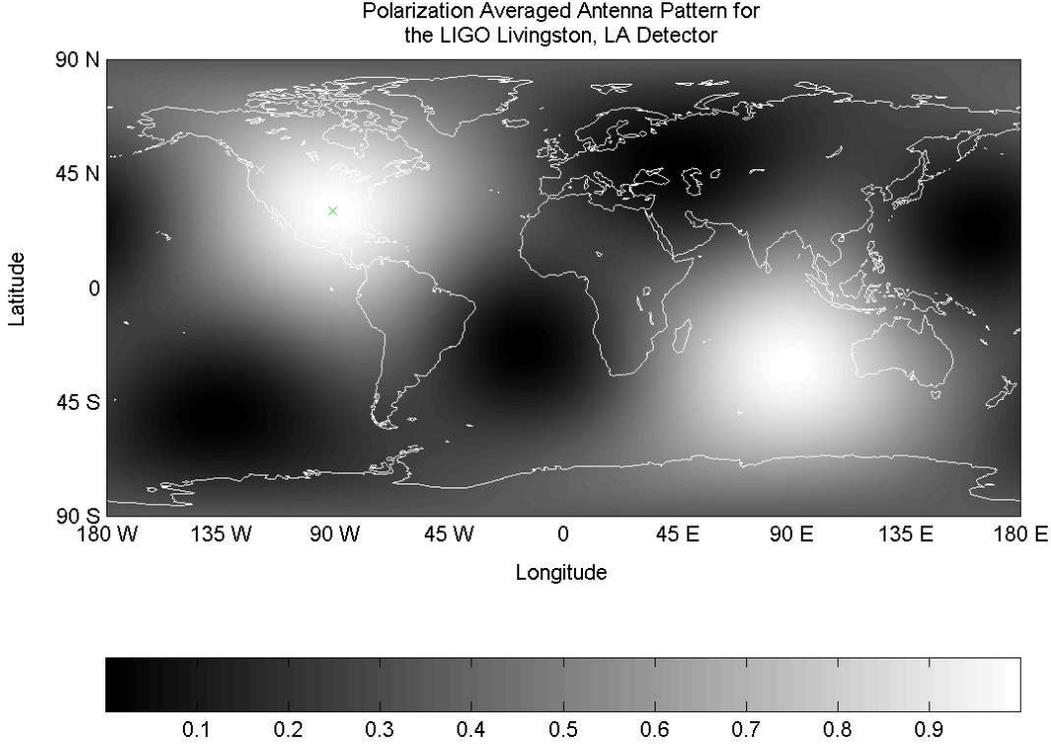}
\caption{Antenna pattern for the LIGO Livingston Observatory, LA}
\label{antennaLLO}
\end{figure}

Example data for the antenna pattern for the LIGO Livingston Observatory, LA is in Figure \ref{antennaLLO}.

The final detector projection, which combines the + and $\times$ polarizations into units of strain as seen by the detector, is then:
\begin{equation}
	h=F_{+}h_{+}+F_{\times}h_{\times}
\end{equation}

The timeseries in units of strain as seen by the detector is output by \texttt{detproj}.

\subsection{Definition of the Polarization Angle $\psi$} \label{psi}

\begin{figure}
\centering
\includegraphics[width=1.0\textwidth]{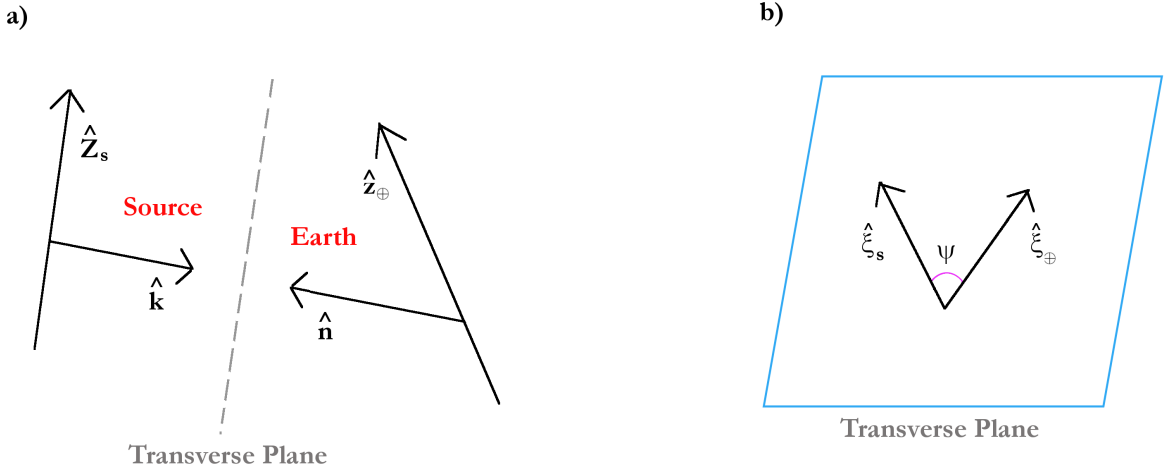}
\caption{a) Definition of the transverse plane between the source and Earth coordinate systems.  b) Definition of the polarization angle $\psi$ as seen on the transverse plane.}
\label{psi3}
\end{figure}

The transverse plane of the TT gauge is defined as a coordinate plane between the Earth and the source that is perpendicular to the line-of-sight.  $\hat{n}$ is defined to be the unit vector pointing from the center of the Earth to the source and $\hat{k}$ is defined to be the unit vector pointing from the center of mass of the source to the Earth, q.v. Figure \ref{psi3}.

Determine the projection of the source's Z-axis and the Earth's Z-axis onto the transverse plane, $\vec{\xi}_{s}$ and $\vec{\xi}_{\oplus}$ respectively:
\begin{eqnarray}
\vec{\xi}_{s}&=&\hat{z}_{s}-\hat{k}(\hat{k}\cdot\hat{z}_{s}) \\
\vec{\xi}_{\oplus}&=&\hat{Z}_{\oplus}-\hat{n}(\hat{n}\cdot\hat{Z}_{\oplus})
\end{eqnarray}

$\psi$ is defined to be the angle between $\vec{\xi}_{s}$ and $\vec{\xi}_{\oplus}$, q.v. Figure \ref{psi3}:

\begin{equation}
\psi=\cos^{-1}\left[\frac{\hat{Z}_{\oplus}-\hat{n}(\hat{n}\cdot\hat{Z}_{\oplus})}{|\hat{Z}_{\oplus}-\hat{n}(\hat{n}\cdot\hat{Z}_{\oplus})|}\cdot\frac{\hat{z}_{s}-\hat{k}(\hat{k}\cdot\hat{z}_{s})}{|\hat{z}_{s}-\hat{k}(\hat{k}\cdot\hat{z}_{s})|}\right]
\end{equation}

Since:
\begin{eqnarray}
\hat{k}\cdot\hat{z}_{s}&=&\cos(\theta_{int}) \\
\hat{n}\cdot\hat{Z}_{\oplus}&=&\cos(\theta_{ext})\\
\hat{k}&=&-\hat{n}
\end{eqnarray}
$\psi$ is then:
\begin{equation}
\psi=\cos^{-1}\left[\frac{\hat{Z}_{\oplus}-\hat{n}\cos(\theta_{ext})}{|\hat{Z}_{\oplus}-\hat{n}\cos(\theta_{ext})|}\cdot\frac{\hat{z}_{s}+\hat{n}\cos(\theta_{int})}{|\hat{z}_{s}+\hat{n}\cos(\theta_{int})|}\right]
\end{equation}

When the Earth's Z-axis is parallel or anti-parallel to the line-of-sight, $\psi$ is measured as the angle between $\hat{X}_{\oplus}$ and $\vec{\xi}_{s}$:
\begin{equation}
\psi=\cos^{-1}\left[\hat{X}_{\oplus}\cdot\frac{\hat{z}_{s}+\hat{n}\cos(\theta_{int})}{|\hat{z}_{s}+\hat{n}\cos(\theta_{int})|}\right]
\end{equation}
When the source's Z-axis is parallel or anti-parallel to the line-of-sight, $\psi$ is measured as the angle between $\vec{\xi}_{\oplus}$ and $\hat{x}_{s}$:
\begin{equation}
\psi=\cos^{-1}\left[\frac{\hat{Z}_{\oplus}-\hat{n}\cos(\theta_{ext})}{|\hat{Z}_{\oplus}-\hat{n}\cos(\theta_{ext})|}\cdot\hat{x}_{s}\right]
\end{equation}
When both the Earth's and the source's Z-axes are parallel or anti-parallel to the line-of-sight, $\psi$ is measured as the angle between $\hat{X}_{\oplus}$ and $\hat{x}_{s}$:
\begin{equation}
\psi=\cos^{-1}\left[\hat{X}_{\oplus}\cdot\hat{x}_{s}\right]
\end{equation}

\section{Injecting Simulated Strain Data Into the Detector Data} \label{calib}

The scope of any gravitational wave simulation package extends to the generation of a strain timeseries as seen by the detector.  The conversion of the strain provided by \texttt{detproj} to the appropriate detector output units, if needed, is an instrumentation issue dealing with the measurement of detector calibration functions.  For detectors whose output is provided in units of strain (e.g. GEO) the simulated strain produced by GravEn may be directly added to the detector data.  

Since LIGO records such things as voltages measured by photodiodes after interference and calibration signals that are injected into the detector hardware separately, calibration information must be used to convert the simulated strain signal into units of detector output before it can be added to the detector data.  While GravEn does include the output of the gravitational wave ``transducer'', i.e., the detector output as a voltage, or in ADC (analog-to-digital-converter) units, the description of this calibration procedure is beyond the scope of this work.  However, appropriate descriptions of the LIGO calibrations is contained in \cite{calibrations} and the use of these to convert strain into LIGO detector output is contained in \cite{GravEnMAN}.

\ack
We are extremely grateful for the help and many contributions, large and small, of John McNabb, Keith Thorne, Shantanu Desai, and Tiffany Summerscales. We are also grateful for the help of Patrick Sutton, especially on matters of calibration.

This work was supported by the Center for Gravitational Wave Physics, and the National Science Foundation under award PHY 00-99559. The Center for Gravitational Wave Physics is supported by the National Science Foundation under cooperative agreement PHY 01- 14375.


\section*{References}
\begin{thebibliography}{1}
\bibitem{MATLAB}
{MATLAB$^{\textregistered}$\ is a product of The MathWorks, Inc.}
\bibitem{GravEnMAN} A.~L. {Stuver}, ``{GravEn Simulation Engine Primer},'' {LIGO}, Tech. Rep. LIGO-T040020-01-Z, 2005.
\bibitem{GravEnCVS} Full source MATLAB$^{\textregistered}$\ code can be found at the following URL: http://www.lsc-group.phys.uwm.edu/cgi-bin/cvs/viewcvs.cgi/matapps/src/simulation/GravEn/?cvsroot=lscsoft
\bibitem{Ottetal} C.~D. {Ott}, A.~{Burrows}, E.~{Livne}, and R.~{Walder}, ``{Gravitational Waves from Axisymmetric, Rotational Stellar Core Collapse},'' \emph{{Astrophys. J.}}, vol. 600, pp. {834--864}, 2004.
\bibitem{Tinto&Schutz} B.~F. {Schutz} and M.~{Tinto}, ``{Antenna patterns of interferometric detectors of gravitational waves. I: Linearly polarized waves},'' \emph{{Mon. Not. R. astr Soc.}}, vol. {224}, pp. {131--154}, {1987}.
\bibitem{300yrs} K. Thorne, ``{Gravitational Radiation}'' in {\underline{300 Years of Gravitation}}, S.~{Hawking} and W.~{Israel}, Eds., pp. 330--458,{New York, NY}: {Cambridge University Press}, 1987.
\bibitem{calibrations} G.~{Gonz\'alez}, M.~{Landry}, B.~{O'Reilly}, and X.~{Siemens}, ``{Calibration of the LIGO detectors for S3},'' LIGO, Tech. Rep. LIGO-T050059-01-D, 2005.
\end{thebibliography}
\end{document}